\documentstyle[aps,prb,multicol,floats,epsf]{revtex}

\begin{document}
\draft
\wideabs{

\title{Zn-doping effect on the magnetotransport properties of \\
Bi$_2$Sr$_{2-x}$La$_x$CuO$_{6+\delta}$ single crystals}

\author{Y. Hanaki$^{1,2,}$\cite{present}, Yoichi Ando$^{1,2,}$\cite{corr}, 
S. Ono$^{1}$, and J. Takeya$^{1}$}

\address{$^1$Central Research Institute of Electric Power Industry, 
Komae, Tokyo 201-8511, Japan}

\address{$^2$Department of Physics, Science University of Tokyo, 
Shinjuku-ku, Tokyo 162-8601, Japan}

\date{\today}
\maketitle

\begin{abstract}
We report the magnetotransport properties of 
Bi$_2$Sr$_{2-x}$La$_x$Cu$_{1-z}$Zn$_z$O$_{6+\delta}$ (Zn-doped BSLCO)
single crystals with $z$ of up to 2.2\%.  
Besides the typical Zn-doping effects on the in-plane resistivity 
and the Hall angle, we demonstrate that the nature of the low-temperature 
normal state in the Zn-doped samples is significantly altered from that 
in the pristine samples under high magnetic fields.
In particular, we observe nearly-isotropic negative magnetoresistance 
as well as an increase in the Hall coefficient at very low temperatures 
in non-superconducting Zn-doped samples, which we propose to be caused 
by the Kondo scattering from the local moments induced by Zn impurities.

\end{abstract}

}
\narrowtext

Effects of nonmagnetic Zn impurities on the electronic properties 
of the high-$T_c$ cuprates have been intensively studied, 
employing almost all available experimental tools.
However, well-controlled studies 
of the Zn-doping effects using high-quality single crystals have 
been mostly limited to the YBa$_2$Cu$_3$O$_{7-\delta}$ (YBCO) system 
and the La$_{2-x}$Sr$_x$CuO$_4$ (LSCO) system because of the availability 
of single crystals; for example, in the otherwise well-studied system of
Bi$_2$Sr$_2$CaCu$_2$O$_{8+\delta}$ (Bi-2212), high-quality single 
crystals can be grown with only up to $\sim$1\% of Zn substitution, 
posing difficulties for systematic studies.
Recently, high-quality single crystals of 
Bi$_2$Sr$_{2-x}$La$_x$CuO$_{6+\delta}$ (BSLCO) have become 
available \cite{Murayama} in a wide range of hole 
concentrations.\cite{Hanaki}  
It is thus natural to investigate the Zn-doping effect in the 
BSLCO system to examine and expand our knowledge of the role of 
Zn impurities in the cuprates.  
Here we report that Zn-doping of up to 2.2\% is possible 
in high-quality BSLCO crystals and present the effect of Zn impurities
on the charge transport properties of this system.

One of the most peculiar charge transport properties of the cuprates 
is that two distinct scattering rates, $\tau_{tr}^{-1}$ and 
$\tau_H^{-1}$, possibly govern the in-plane resistivity 
$\rho_{ab}$ and the Hall angle $\theta_H$, respectively.
Zn impurities have been believed to induce \cite{Chien,Fukuzumi} 
residual terms in both $\tau_{tr}^{-1}$ and $\tau_H^{-1}$, 
which ultimately lead to charge localization \cite{Segawa};
these residual scattering rates may well be related \cite{Nagaosa} to 
the local moments induced by the nonmagnetic Zn impurities, \cite{Alloul} 
though the role of the local moments in the charge transport in cuprates 
is not well understood yet.
It would thus be meaningful to look for some peculiar features that is 
clearly due to the local moments in the transport properties of the 
Zn-doped BSLCO crystals.

Perhaps the best-known effect of Zn-doping in cuprates is the rapid 
suppression of $T_c$.  The rate of $T_c$ suppression has been known 
to be around 10--15 K/at.\% upon Zn substitution.  It has been suggested 
\cite{Koike,Akoshima} that this rate can be enhanced to $\sim$20 K/at.\% 
near the hole concentration per Cu, $p$, of 1/8, 
which was discussed to be due to the ``pinning" \cite{Hirota} 
of the charge stripes \cite{Tranquada} by Zn impurities. 
Since it was recently revealed \cite{Ono} 
that the metal-to-insulator (M-I) crossover 
in the low-temperature normal state of BSLCO occurs at $p \simeq 1/8$, 
it is particularly intriguing to look at the effect 
of Zn at $p \simeq 1/8$ to elucidate the impact of the 
charge-stripes instability in BSLCO.

In this paper, we report the charge transport properties in 
Zn-doped BSLCO (Bi$_2$Sr$_{2-x}$La$_x$Cu$_{1-z}$Zn$_z$O$_{6+\delta}$) 
single crystals with $z$ of up to 2.2\% for $x$ = 0.50 and 0.66.
These La contents correspond to $p$ of $\sim$0.15 and $\sim$1/8, 
respectively,\cite{Hanaki} and 2.2\%-Zn is enough to completely 
suppress superconductivity at $x$ = 0.66.
We observe more or less standard Zn-doping effects 
in the in-plane resistivity, Hall angle, and $T_c$,
without any noticeable ``1/8 anomaly" in the Zn-doped samples.
On the other hand, we found very peculiar negative magnetoresistance 
as well as an upturn in $R_H$ at low temperatures in non-superconducting 
samples, which is most likely to be related to the 
local moments induced by Zn.

The crystals are grown by a floating-zone technique as 
reported previously.\cite{Murayama}  
We have demonstrated \cite{Murayama,Hanaki,Ono} that our crystals are 
among the best available BSLCO crystals in terms of 
optimum $T_c$ (which is as high as 38 K), residual resistivity, 
and the control of hole doping. 
The La concentrations in the crystals are determined by the 
electron-probe microanalysis (EPMA), and the actual Zn concentrations 
are measured by the inductively-coupled plasma (ICP) analysis; the 
errors in $x$ and $z$ are estimated to be $\pm 0.01$ and $\pm 0.002$, 
respectively. 
For the transport measurements, the crystals are cut into dimensions 
typically 1 $\times$ 0.5 $\times$ 0.02 mm$^3$. 
The thickness of the samples are calculated from their weight (measured 
with 0.1-$\mu$g resolution) to accurately determine the absolute values 
of $\rho_{ab}$ and the Hall coefficient $R_H$.
All the crystals are annealed in air at 650$^{\circ}$C for 48 h and 
quenched to room temperature to achieve uniform oxygen distribution. 
A standard ac six-probe method is employed to measure $\rho_{ab}$ and $R_H$.
The magnetoresistance and $R_H$ are measured by sweeping the magnetic 
field to both plus and minus polarities at constant 
temperatures.\cite{Murayama}  
For the non-superconducting samples with ($x$,$z$) = (0.66,0.022), 
we measure both the transverse and longitudinal magnetoresistance (MR) down 
to 450 mK in a $^3$He refrigerator where the temperature control during 
the magnetic-field sweep is done with a stability of $\sim$1 mK using a 
capacitance sensor embedded in the sample stage.

\begin{figure}[t!]
\epsfxsize=0.85\columnwidth
\centerline{\epsffile{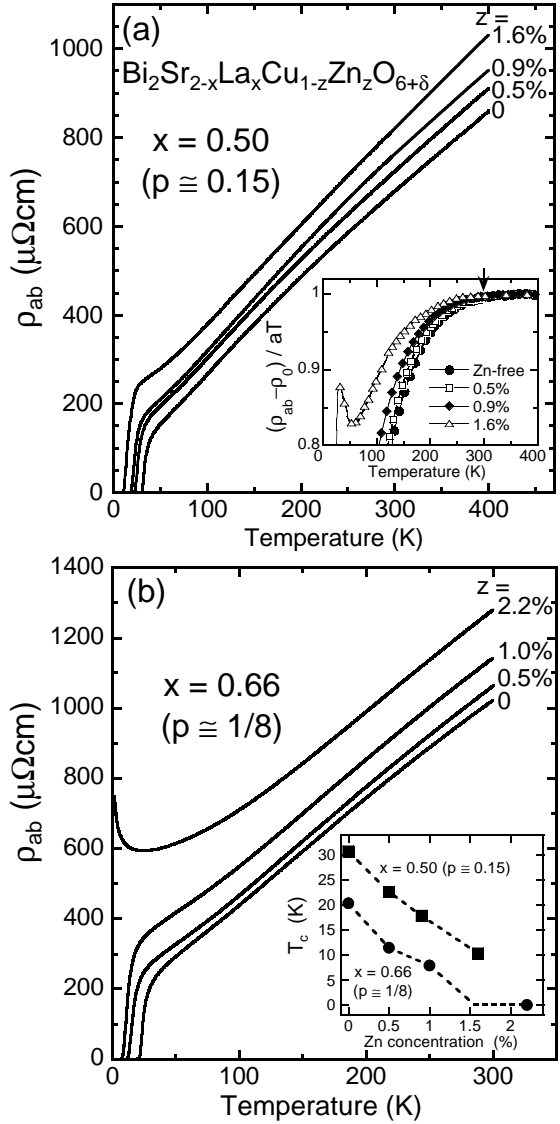}}
\vspace{0.2cm}
\caption{$\rho_{ab}(T)$ data of 
Bi$_2$Sr$_{2-x}$La$_x$Cu$_{1-z}$Zn$_z$O$_{6+\delta}$ single crystals 
at (a) $x$ = 0.50 and (b) $x$ = 0.66 with various $z$. 
Inset to (a): Plots of $(\rho_{ab}-\rho_{0})/aT$ vs. $T$ for 
$x$ = 0.50 with the four $z$ values; arrow marks $T^*$.
Inset to (b): Plots of $T_c$ vs. $z$ for $x$ = 0.50 and 0.66.}
\label{fig1}
\end{figure}

Figures 1(a) and 1(b) show temperature dependences of $\rho_{ab}$ for 
$x$ = 0.50 and 0.66 with various Zn concentrations. 
One can see that at both $x$ the $\rho_{ab}(T)$ curves are almost 
parallel-shifted upon Zn doping, indicating that the main effect of 
Zn impurities on $\rho_{ab}$ is to increase a temperature-independent 
residual term in $\tau_{tr}^{-1}$.
In the $x$ = 0.50 series [Fig. 1(a)], there is a reasonably wide region 
of $T$-linear resistivity at high temperatures, so that we can 
estimate the pseudogap temperature $T^*$ for each Zn concentration 
from the downward deviation of the $\rho_{ab}(T)$ curves from the 
$T$-linear behavior.  The inset of Fig. 1(a) shows the plots of 
$[\rho_{ab}(T)-\rho_{0}]/aT$ vs. $T$ (where $a$ is the 
$T$-linear slope and $\rho_0$ is the zero-temperature intercept of the 
$T$-linear behavior), which make it clear that the deviation takes 
place at nearly the same temperature ($\sim$300 K) for all $z$. 
Thus, as was reported \cite{Mizuhashi} for YBCO, $T^*$ as determined 
from $\rho_{ab}(T)$ does not move with Zn doping in BSLCO.  

In the $x$ = 0.66 series [Fig. 1(b)], it is notable that the 
superconductor-to-insulator (S-I) transition occurs around 
$\rho_{ab}$ of $\sim$400 $\mu\Omega$cm, which corresponds to the 
sheet resistance per CuO$_2$ plane of $\sim$3.3 k$\Omega$;
this is half the quantum value $h/(2e)^2$ ($\simeq$6.5 k$\Omega$)
and thus differs by a factor of 2 from the result of the 
critical sheet resistance obtained \cite{Fukuzumi,Mizuhashi} for YBCO 
and LSCO, indicating that the ``universal" critical sheet resistance 
for the S-I transition is not exactly universal for the cuprates.
We emphasize that the uncertainty in the absolute value of $\rho_{ab}$ 
is less than 10\% in our measurements.\cite{Ono}

The inset of Fig. 1(b) shows the suppression of $T_c$ upon Zn doping 
for the two $x$ values; the suppression rates are almost the same for 
the two cases and is about 13 K/at.\%, which is typical for the cuprates.
Note that there is no enhancement in the $T_c$-suppression rate for 
$p \simeq 1/8$ ($x$ = 0.66), and thus the sort of amplification of the 
1/8 anomaly suggested \cite{Koike,Akoshima} for LSCO and Bi-2212 
are not observed in BSLCO.  
This can be interpreted to mean that the charge-order instability 
at $p \simeq 1/8$ is so weak \cite{note1} in BSLCO that the 
``pinning" by the Zn impurities are not effective, 
though this interpretation is highly speculative.

\begin{figure}[t!]
\epsfxsize=1.0\columnwidth
\centerline{\epsffile{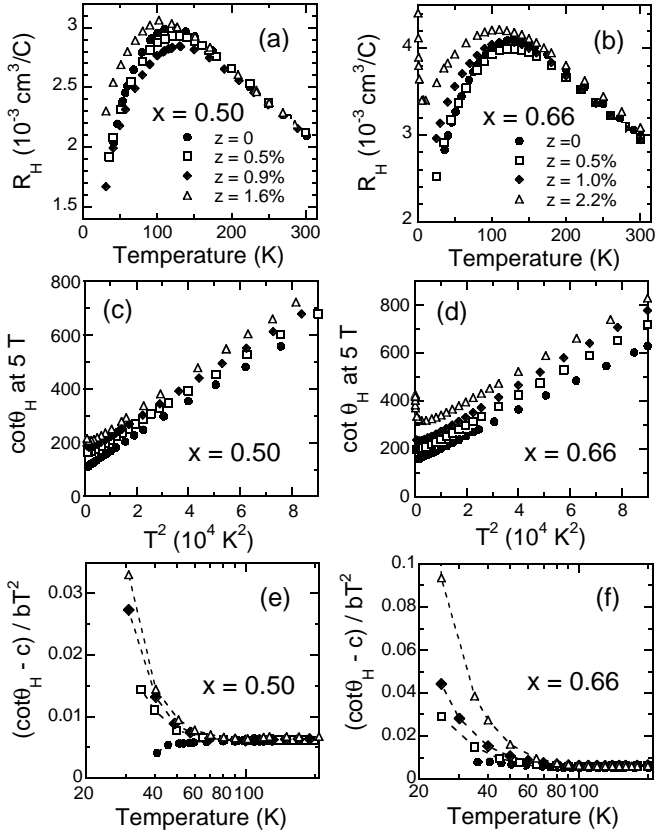}}
\vspace{0.2cm}
\caption{(a),(b): $T$ dependences of $R_H$ for (a) $x$ = 0.50 and 
(b) $x$ = 0.66 with various $z$.
(c),(d): Plots of $\cot \theta _H$ vs. $T^2$ for the two series.
(e),(f): Plots of $(\cot \theta_H - c)/bT^2$ vs. $T$ for the two series, 
which emphasize the deviation from the $T^2$ behavior.}
\label{fig2}
\end{figure}

Figures 2(a) and 2(b) show the temperature dependences of $R_H$ for 
the two series.  For each $x$, the magnitude of $R_H$ does not change 
with $z$ above 200 K, demonstrating that the Zn substitution does 
not change the hole concentration. 
At lower temperatures, $R_H$ becomes $z$ dependent and the peak in 
$R_H(T)$ shows a non-monotonic change upon Zn doping.
As is the case with other systems, the complicated change in 
$R_H(T)$ can be simplified by looking at the Hall angle.
Figures 2(c) and 2(d) show the plots of $\cot \theta _H$ vs. $T^2$ for 
the two series.  All the data of $\cot \theta_H$ are almost linear in 
$T^2$ and appear to be parallel-shifted upon Zn doping, suggesting that 
the Zn impurities increase a temperature-independent 
residual term in $\tau_{H}^{-1}$; this is actually the behavior that 
led to the two-scattering-rate scenario \cite{Chien} and thus is 
typical for the cuprates. 
Upon closer look at the data in Figs. 2(c) and 2(d), one may notice 
that at low temperatures there is an upward deviation from the $T^2$ 
behavior in the Zn-doped samples.  
To make this point clear, Figs. 2(e) and 2(f) show the plots of 
$(\cot \theta_H - c)/bT^2$ vs. $T$, where $b$ is the $T^2$ slope and 
$c$ is the zero-temperature intercept of the linear-in-$T^2$ behavior. 
One can see that the deviation becomes systematically more pronounced 
as $z$ is increased; this behavior most likely reflects some 
localization effect in the Hall channel and is probably responsible 
for the weakening of the temperature dependence of $R_H$ upon Zn doping.

\begin{figure}[t!]
\epsfxsize=0.9\columnwidth
\centerline{\epsffile{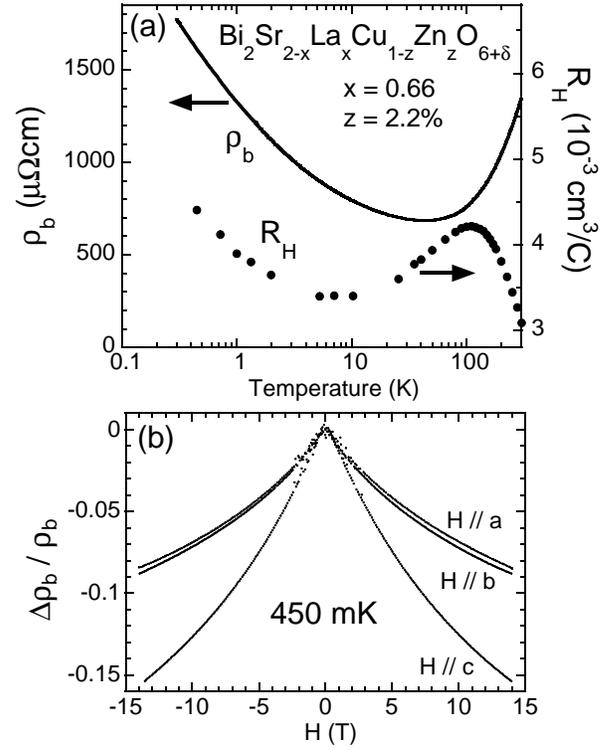}}
\vspace{0.2cm}
\caption{(a) $\log T$ plot of $\rho_{b}$ of a 2.2\%-Zn-doped 
$x$ = 0.66 sample; $R_H(T)$ is also plotted using the right-hand-side 
axis. 
(b) Magnetoresistance of the above $\rho_b$-sample ($I \parallel b$) 
at 450 mK for $H$ along $a$, $b$, and $c$.}
\label{fig3}
\end{figure}

As already noted, the superconductivity is completely suppressed 
in the 2.2\%-Zn-doped samples at $x$ = 0.66, in which we can measure 
the normal-state transport properties down to very low temperatures. 
Incidentally, the low-temperature normal-state transport has already 
been measured for pristine BSLCO crystals with $x$ = 0.66 under 60-T 
magnetic field,\cite{Ono} so we can directly compare the normal state 
brought about by Zn-doping to that brought about by high magnetic fields.
As is shown in Fig. 3(a),\cite{note} 
$\rho_{ab}(T)$ increases with decreasing temperature below $\sim$30 K 
and the temperature dependence is quicker than $\log (1/T)$, 
which is inferred from a positive curvature in this semi-log plot; 
this behavior is contrasting to the ``metallic" behavior 
found \cite{Ono} in the pristine sample with $x$ = 0.66, 
where $\rho_{ab}$ becomes 
temperature independent below $\sim$10 K with $\rho_{ab} \simeq 190$ 
$\mu\Omega$cm.  Therefore, the localization behavior in the 
Zn-doped sample is clearly due to the additional scattering caused by 
the Zn impurities.  Also, since the ``insulating" behavior found \cite{Ono} 
in more underdoped pristine samples ($p < 1/8$) is consistent with 
$\log(1/T)$ (which is slower than the behavior of the Zn-doped samples), 
the nature of the charge-localized state in the Zn-doped sample is 
apparently different from that in the pristine sample under high magnetic 
field.  

It is found that $R_H$ of the non-superconducting Zn-doped samples 
also shows an upturn at low temperatures [Fig. 3(a)]; this is again 
in contrast to the $R_H(T)$ behavior of the pristine samples 
\cite{Ando_Hall} under 60 T, which becomes essentially temperature 
independent at low temperatures. The rather strong temperature dependence 
of $R_H$ below $\sim$3 K in the Zn-doped samples strongly suggests 
that the Zn-induced charge localization is {\it not} due to a 
simple weak localization effect.\cite{Ando_Hall}

We can obtain further insight into the Zn-induced localized state 
from the magnetoresistance (MR), which turns out to be negative at 
low temperatures and is very peculiar.  
Figure 3(b) shows the MR at 450 mK for three geometries; the sample 
was cut so that the current $I$ flows along the $b$-axis, and the 
magnetic field $H$ is applied along $a$, $b$ (longitudinal geometry),
and $c$ (transverse geometry).  There is essentially no anisotropy 
between $H \parallel a$ and $H \parallel b$.  
The anisotropy between $H \parallel b$ and $H \parallel c$ is 
less than a factor of 2 and the $H$ dependences for the two geometries 
are almost exactly the same.  
Therefore, we can conclude that the MR is essentially isotropic 
and thus is of spin origin.
The $H$ dependence of this negative MR is not $H^2$ but is reminiscent 
of the $H$ dependence expected for Kondo 
scattering.\cite{Andrei,Schlottmann,Ruvalds} 
Since the Kondo effect should yield isotropic MR, the main features of 
the MR observed here appears to be consistent, at least qualitatively, 
with what is expected for the Kondo effect.
It is useful to note that a very anisotropic negative MR coming from 
weak localization has been observed \cite{Jing} in non-superconducting 
samples of pristine Bi$_2$Sr$_2$CuO$_{6+\delta}$, which is contrasting 
to our observation in the Zn-doped samples.

\begin{figure}[t!]
\epsfxsize=1.0\columnwidth
\centerline{\epsffile{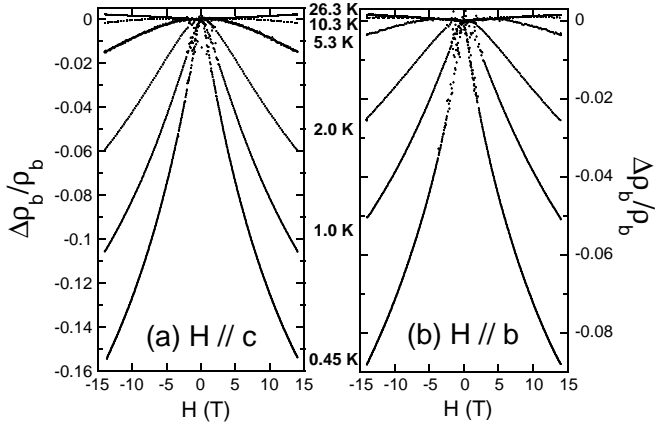}}
\vspace{0.2cm}
\caption{Magnetoresistance of a 2.2\%-Zn-doped $x$ = 0.66 sample for 
(a) $H \parallel c$ (transverse) and (b) $H \parallel b$ (longitudinal) 
at low temperatures ($I \parallel b$).}
\label{fig4}
\end{figure}

The evolution of the MR with temperature is shown in Fig. 4 for 
$H \parallel c$ and $H \parallel b$.  
The MR becomes predominantly negative below $\sim$5 K, which is 
nearly the same as the onset temperature of the low-temperature upturn 
in $R_H(T)$, and therefore the negative MR and the increase 
in $R_H$ appear to have a common origin.  
It is useful to note that the Kondo scattering is expected to become 
effective simultaneously in the MR and in the Hall effect.\cite{Ruvalds}
Thus, all the above results of the low-temperature MR and $R_H$ seem to 
be most consistent with the Kondo effect, which may be caused 
\cite{Nagaosa,Alloul2,Rullier} by the local moments \cite{Alloul} induced 
by the Zn impurities.
The possibility that the Kondo effect is playing a major role in the 
Zn-doped samples also explains the various differences 
between the normal state in the Zn-doped samples and that in pristine 
samples under high magnetic fields. 
For more quantitative understanding of the possible Kondo effect in 
the cuprates, theoretical calculations for the Kondo scattering in the 
non-Fermi-liquid ground state of the cuprates would be required.

To summarize, in the BSLCO crystals we observed quite typical 
Zn-doping effects on $\rho_{ab}(T)$ 
and $\cot \theta_H(T)$, and found no 1/8 anomaly in the $T_c$ suppression 
rate.  Comparison of the low-temperature normal state brought about by 
Zn-doping to that brought about by high magnetic field reveals significant 
difference between the two, which highlights the peculiar nature of 
the charge-localized state in the Zn-doped samples.  The negative MR 
and the upturn in $R_H$ observed at low temperatures strongly suggest 
that the Kondo scattering due to the local moments induced by Zn is 
playing a key role.

We thank S. Uchida and A. N. Lavrov for helpful discussions and 
T. Murayama for technical assistance.

%
\medskip
\vfil


\begin{references}
\bibitem{note1}
This observation is actually in good accord with the M-I crossover 
\cite{Ono} occurring at $p \simeq 1/8$ in pristine BSLCO, 
because such M-I crossover could mean that the static stripes 
(which presumably cause charge localization in the normal state) 
disappear above $p \simeq 1/8$ in BSLCO.
\bibitem{note}
The sample measured for Fig. 3(a) is different from that for Fig. 1(b).  
We have measured more than 10 samples from the same batch of 
($x$,$z$) = (0.66, 0.022), and found that the resistivity values at 
300 K differ by up to 20\%, which is presumably due to slight differences 
in the Zn concentration.  The low-temperature localization behavior is 
found to be correspondingly different, becoming stronger as 
$\rho_{ab}$ at 300 K becomes larger.  However, the MR behavior 
is essentially reproducible in all the samples measured.
\end{references}
\end{document}